\documentstyle[aps,eqsecnum,epsfig]{revtex}

\begin{document}
\draft
\title{Radiative Corrections to the Single Spin Asymmetry\\[0pt]
in Heavy Quark Photoproduction\footnote{Contribution to XV 
International Seminar on High Energy Physics Problems,  Dubna, Russia,  Sept. 25-29, 2000}}
\author{\large N.Ya.Ivanov\footnote{E-mail: nikiv@uniphi.yerphi.am}}
\address{Yerevan Physics Institute, Alikhanian Br.2, 375036 Yerevan,
Armenia\\}
\maketitle
\begin{abstract}
\noindent We analyze in the framework of pQCD the properties of the single
spin asymmetry in heavy flavor production by linearly polarized photons. At
leading order, the parallel- perpendicular asymmetry in azimuthal
distributions of both charm and bottom quark is predicted to be about $20\%$
in a wide region of initial energy. Using the soft gluon resummation formalism,
we have calculated the next-to-leading order corrections to the asymmetry to 
next-to-leading logarithmic accuracy. It is shown that radiative corrections practically 
do not affect the Born predictions for the azimuthal asymmetry at energies of the fixed 
target experiments. Both leading and next-to-leading order predictions for the
asymmetry are insensitive to within few percent to theoretical uncertainties
in the QCD input parameters: $m_{Q}$, $\mu _{R}$, $\mu _{F}$, $\Lambda _{QCD}
$ and in the gluon distribution function.  Our analysis shows that nonperturbative 
corrections to a $B$-meson azimuthal asymmetry due to the gluon transverse motion
in the target are negligible. We conclude that measurements of the single spin 
asymmetry would provide a good test of pQCD applicability to heavy flavor 
production at fixed target energies.\\[0pt]
\end{abstract}

\section{ Introduction}

\noindent Presently, the basic spin-averaged characteristics of heavy flavor hadro-, 
photo- and electroproduction are known exactly up to the next-to-leading 
order (NLO) \cite{1,2,3,4,5,6,7,8,9}. Two main results of  the exact pQCD 
calculations can be formulated as follows.
First, the NLO corrections are large; they increase the leading order
(LO) predictions for both charm and bottom production cross sections
approximately by a factor of 2. For this reason, one could expect that the
higher order corrections as well as the nonperturbative contributions can be
essential in these processes, especially for the $c $-quark case.
Second, the fixed order predictions are very sensitive to standard uncertainties
in the input QCD parameters. In fact, the total uncertainties associated with 
the unknown values of the heavy quark mass, $m_{Q}$, the factorization 
and renormalization scales, $\mu _{F}$ and $\mu _{R}$, $\Lambda _{QCD}$ 
and the parton distribution functions are so large that one can only estimate 
the order of magnitude of the NLO predictions for total cross sections \cite{7,8}.
For this reason,  it is very difficult to compare directly, without additional
assumptions, the fixed order predictions for spin-averaged cross sections with
experimental data and thereby to test the pQCD applicability  to the heavy 
quark production.

Since the spin-averaged characteristics of heavy flavor production are not
well defined quantitatively in pQCD it is of special interest to study those
spin-dependent observables which are stable under variations of input 
parameters of the theory \cite{we}. In this report we analyze the charm and 
bottom production by linearly polarized photons, namely the reactions
\begin{equation}
\gamma \uparrow +N\rightarrow Q(\overline{Q})+X.  \label{1}
\end{equation}
We consider the single spin asymmetry parameter, $A(s)$, which measures 
the parallel-perpendicular asymmetry in the quark azimuthal distribution:
\begin{equation}
A(s)=\frac{1}{{\cal P}_{\gamma }}\frac{\mbox{d}\sigma (s,\varphi =0)-\mbox{d}
\sigma (s,\varphi =\pi /2)}{\mbox{d}\sigma (s,\varphi =0)+\mbox{d}\sigma
(s,\varphi =\pi /2)}.  \label{2}
\end{equation}
Here d$\sigma (s,\varphi )\equiv \frac{\mbox{d}\sigma }{\mbox{d}\varphi }
(s,\varphi )$ , ${\cal P}_{\gamma }$ is the degree of linear polarization of
the incident photon beam, $\sqrt{s}$ is the centre of mass energy of the
process (\ref{1}) and $\varphi $ is the angle between the beam polarization
direction and the observed quark transverse momentum. 

The properties of the single spin asymmetry at Born level as well as the 
contributions of  nonperturbative effects (such as the gluon transverse 
motion in the target and the heavy quark fragmentation) have been considered
in \cite{we}. Using the methods of Refs.\cite{r1,15,16,17} for the threshold 
resummation of soft gluons,  we have also calculated the next-to-leading 
order corrections to $A(s)$ at next-to-leading logarithmic (NLL) level \cite{we2}.
The main results of our analysis can be formulated as follows:
\begin{itemize}
\item  At fixed target energies, the LO predictions for azimuthal asymmetry (\ref
{2}) are not small and can be tested experimentally. For instance,
\begin{equation}
A(s=400\rm{ GeV}^{2})\mid _{\rm{Charm}}^{\rm{LO}}\approx A(s=400\rm{
GeV}^{2})\mid _{\rm{Bottom}}^{\rm{LO}}\approx 0.18.  \label{1AN}
\end{equation}
\item Radiative corrections practically do not affect the
Born predictions for $A(s)$ at fixed target energies.
\item At energies sufficiently above the production threshold, both leading and
next-to-leading order predictions for $A(s)$ are insensitive (to within few
percent) to uncertainties in the QCD parameters: $m_{Q}$, $\mu _{R}$, $\mu
_{F}$, $\Lambda _{QCD}$ and in the gluon distribution function. This implies
that theoretical uncertainties in the spin-dependent and spin-averaged cross
sections (the numerator and denominator of the fraction (\ref{2}),
respectively) cancel each other with a good accuracy.
\item Nonperturbative corrections to the $b$-quark azimuthal asymmetry 
$A(s)$ due to the gluon transverse motion in the target are negligible. 
Because of the smallness of the $c$-quark mass, the $k_{T}$-kick corrections 
to $A(s)$ in the charm case are larger; they are of order of 20\%.
\end{itemize}

We conclude that the single spin asymmetry is an observable quantitatively
well defined, rapidly convergent in pQCD and  insensitive to nonperturbative 
contributions. Measurements of asymmetry parameters would provide a good 
test of the fixed order QCD applicability to heavy flavor production.

\section{Single spin asymmetry at leading order}

\noindent At leading order, ${\cal O}(\alpha _{em}\alpha _{S})$, the only partonic
subprocess which is responsible for heavy quark photoproduction is the
two-body photon-gluon fusion:
\begin{equation}
\gamma (k_{\gamma })+g(k_{g})\rightarrow Q(p_{Q})+\overline{Q}(p_{\stackrel{
\_}{Q}}).  \label{3}
\end{equation}
The cross section corresponding to the Born diagrams is \cite{we}:
\begin{equation}
\frac{\mbox{d}^{2}\hat{\sigma}^{\rm{Born}}}{\mbox{d}\hat{x}\mbox{d}\varphi }( \hat{s},%
\hat{t},\varphi ,\mu _{R}^{2})=C\frac{e_{Q}^{2}\alpha _{em}\alpha _{S}(\mu
_{R}^{2})}{\hat{s}}\left[ \frac{1+\hat{x}^{2}}{ 1-\hat{x}^{2}}+\frac{%
2(1-\beta ^{2})(\beta ^{2}-\hat{x}^{2})}{(1- \hat{x}^{2})^{2}}\left( 1+{\cal %
P}_{\gamma }\cos 2\varphi \right) \right],  \label{4}
\end{equation}
where ${\cal P}_{\gamma }$ is the degree of the photon beam polarization; $%
\varphi $ is the angle between the observed quark transverse momentum, $\vec{%
p}_{Q,T}$, and the beam polarization direction. In (\ref{4}) $C$ is the
color factor, $C=T_{F}=$Tr$(T^{a}T^{a})/(N_{c}^{2}-1)=1/2$, and $e_{Q}$ is
the quark charge in units of electromagnetic coupling constant. We use the
following definition of partonic kinematical variables: 
\begin{eqnarray}
\hat{s} &=&\left( k_{\gamma }+k_{g}\right) ^{2};\mbox{ \ \hspace{0.1in} ~ 
\hspace{0.1in} \hspace{0.05in}}\hat{t}=\left( k_{\gamma }-p_{Q}\right) ^{2};
\nonumber \\
\hat{u} &=&\left( k_{g}-p_{Q}\right) ^{2};\mbox{ \hspace{0.21in} \hspace{
0.09in} \hspace{0.05in}}\hat{x}=1+2\frac{\hat{t}-m^{2}}{\hat{ s}};
\label{5} \\
\beta &=&\sqrt{1-\frac{4m^{2}}{\hat{s}}};\mbox{ \hspace{0.24in} \hspace{%
0.08in} }\vec{p}_{Q,T}^{\hspace{0.02in}2}=\frac{\hat{s}}{4}\left( \beta ^{2}-%
\hat{x}^{2}\right) ;  \nonumber
\end{eqnarray}
where $m$ is the heavy quark mass.

Unless otherwise stated, the CTEQ5 \cite{11} parametrization of the gluon
distribution function is used. The default values of the charm and bottom
mass are $m_{c}=$ 1.5 GeV and $m_{b}=$ 4.75 GeV; $\Lambda _{4}=$ 300 MeV and 
$\Lambda _{5}=$ 200 MeV. The default values of the factorization scale $\mu
_{F}$ chosen for the $A(s)$ asymmetry calculation are $\mu _{F}\mid _{\rm{
Charm}}=2m_{c}$ for the case of charm production and $\mu _{F}\mid _{\rm{
Bottom}}=m_{b}$ for the bottom case \cite{9,10}. For the renormalization
scale, $\mu _{R}$, we use $\mu _{R}=\mu _{F}$.

Let us discuss the pQCD predictions for the asymmetry parameter defined by (%
\ref{2}). Our calculations of $A(s)$ at LO for the $c$- and $b$-quark are
presented by solid lines in Fig.1. One can see that
at energies sufficiently above the production threshold the single spin
asymmetry $A(E_{\gamma })$ depends weekly on $E_{\gamma }$, $%
E_{\gamma}=(s-m_{N}^2)/2m_{N}$.
\begin{figure}
\begin{center}
\begin{tabular}{cc}
\mbox{\epsfig{file=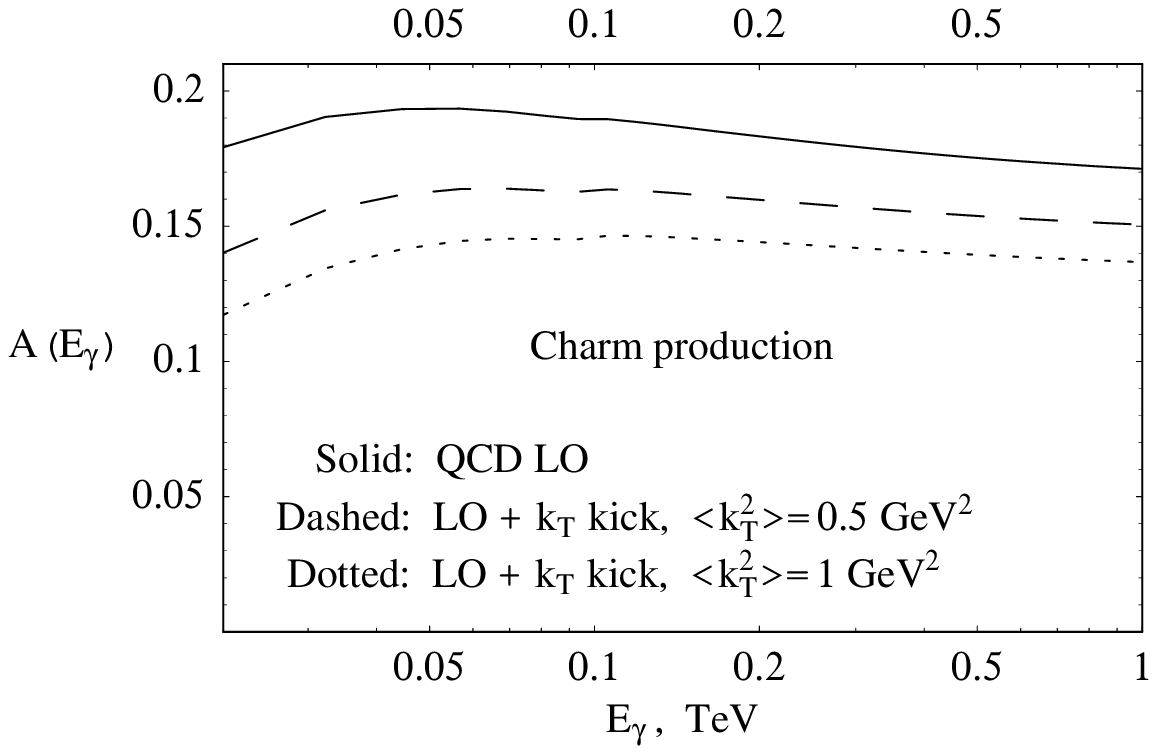,width=250pt}}
&\mbox{\epsfig{file=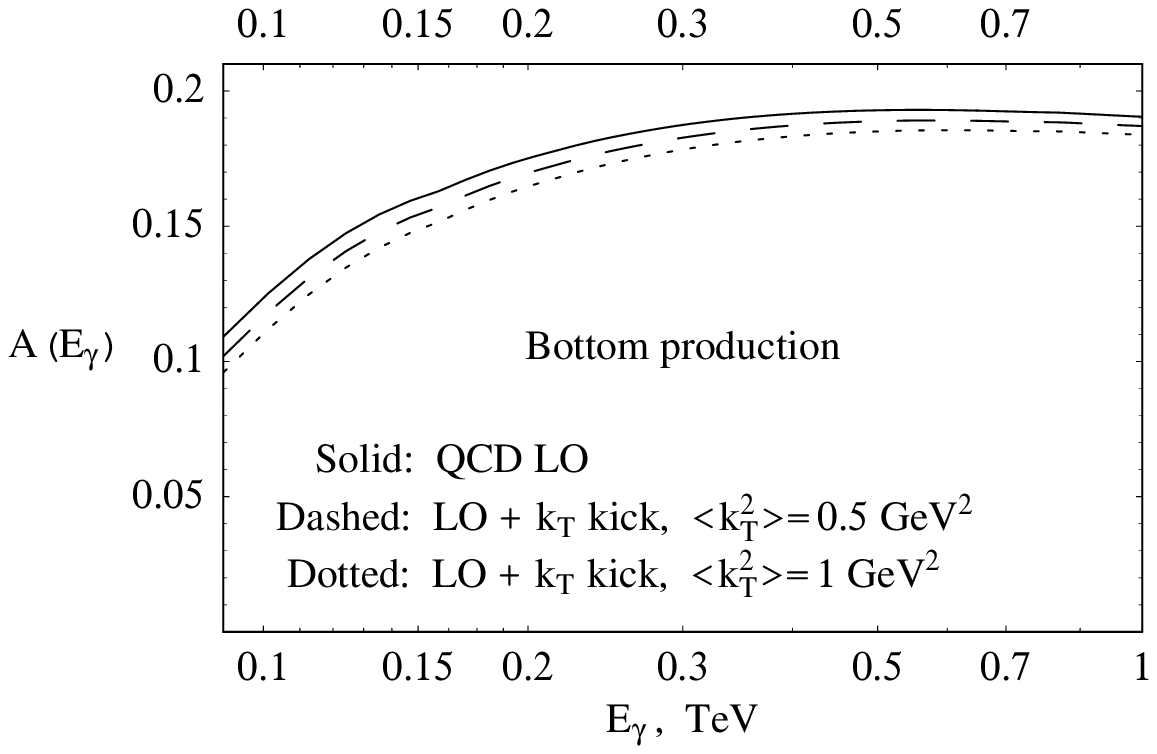,width=250pt}}\\
\end{tabular}
\caption{\small Single spin asymmetry, $A(E_{\gamma})$, in  $c$- and  $b$- quark
production as a function of beam energy $E_{\gamma}=(s-m_{N}^2)/2m_{N}$;
the QCD LO predictions with and without the inclusion of $k_{T}$-kick effect.}
\end{center}
\end{figure}

The most interesting feature of LO predictions for $A(E_{\gamma })$ is that
they are practically insensitive to uncertainties in QCD parameters. In
particular, changes of the charm quark mass in the interval 1.2 $<m_{c}<$
1.8 GeV affect the quantity $A(E_{\gamma })$ by less than 6\% at energies 40 
$<E_{\gamma }<$ 1000 GeV. Remember that analogous changes of $m_{c}$ lead to
variations of total cross sections from a factor of 10 at $E_{\gamma }=$ 40
GeV to a factor of 3 at $E_{\gamma }=$ 1 TeV. The extreme choices $m_{b}=$
4.5 and $m_{b}=$ 5 GeV lead to 3\% variations of the parameter $A(E_{\gamma
})$ in the case of bottom production at energies 250 $<E_{\gamma }<$ 1000
GeV. The total cross sections in this case vary from a factor of 3 at $%
E_{\gamma }=$ 250 GeV to a factor of 1.5 at $E_{\gamma }=$ 1 TeV. The
changes of $A(E_{\gamma })$ are less than 3\% for choices of $\mu _{F}$ in
the range $\frac{1}{2}m_{b}<\mu _{F}<2m_{b}$. For the total cross sections,
such changes of $\mu _{F}$ lead to a factor of 2.7 at $E_{\gamma }=$ 250 GeV
and of 1.7 at $E_{\gamma }=$ 1 TeV. We have verified also that all the CTEQ3-CTEQ5
versions of the gluon distribution function \cite{11} as well as the CMKT
parametrization \cite{12} lead to asymmetry predictions which coincide with
each other with accuracy better than 1.5\%

\section{Nonperturbative contributions}

\noindent Let us discuss how the pQCD predictions for single spin asymmetry are
affected by nonperturbative contributions due to the intrinsic transverse
motion of the gluon in the target.  In our analysis, we use the MNR model 
\cite{10} parametrization of the gluon transverse momentum distribution,
\begin{equation}
\vec{k}_{g}=z\vec{k}_{N}+\vec{k}_{g,T},  \label{3AN}
\end{equation}
where $\vec{k}_{N}$ is the target momentum in the $\gamma N$ centre of mass system.
According to \cite{10}, the primordial transverse momentum, $\vec{ k}_{g,T}$
, has a random Gaussian distribution:
\begin{equation}
\frac{1}{N}\frac{\mbox{d}^{2}N}{\mbox{d}^{2}k_{T}}=\frac{1}{\pi \langle
k_{T}^{2}\rangle }\exp \left( -\frac{k_{T}^{2}}{\langle k_{T}^{2}\rangle }
\right),  \label{12}
\end{equation}
where $k_{T}^{2}\equiv \vec{k}_{g,T}^{\hspace{0.02in}2}$. It is evident that
the inclusion of this effect results in a dilution of azimuthal asymmetry.
In \cite{9,10}, the parametrization (\ref{12}) (so-called $k_{T}$-kick) 
have been used to describe the single inclusive spectra and the $Q \overline{
Q}$ correlations. It was found that in charm photoproduction 0.5 $< \langle
k_{T}^{2}\rangle <$ 2 GeV$^{\rm{2}}$.

Our calculations of the parameter $A(s)$ at LO with the $k_{T}$-kick
contributions are presented in Fig.1  by dashed ($\langle
k_{T}^{2}\rangle =$ 0.5 GeV$^{\rm{2}}$) and dotted ($\langle
k_{T}^{2}\rangle =$ 1 GeV$^{\rm{2}}$) curves.

So, we can conclude that nonperturbative corrections to the $b$-quark
asymmetry parameter (\ref{2}) due to the $k_{T}$-kick effect practically do
not affect predictions of the underlying perturbative mechanism:
photon-gluon fusion.

Our calculations of the $p_{T}$- and $x_{F}$
-distributions of the azimuthal asymmetry in heavy quark photoproduction are
given in \cite{we}.

\section{Next-to-leading order corrections}

\noindent The perturbative expansion for the $\gamma g$ cross section, $\hat{\sigma}
(\rho ,\mu ^{2})$, is usually written in terms of scaling functions:
\begin{equation}
\hat{\sigma}(\rho ,\mu ^{2})=\frac{\alpha _{em}\alpha _{S}(\mu ^{2})e_{Q}^{2}
}{m^{2}}\sum_{k=0}^{\infty }\left( 4\pi \alpha _{S}(\mu ^{2})\right)
^{k}\sum_{l=0}^{k}c^{(k,l)}(\rho )\ln ^{l}\frac{\mu ^{2}}{m^{2}},
\label{4.2}
\end{equation}
where $\rho =4m^{2}/\hat{s}$, $\hat{s}=\left( k_{\gamma }+k_{g}\right) ^{2}$
; $c^{(0,0)}(\rho )$ corresponds for the Born contribution (\ref{4}), $
c^{(1,1)}(\rho )$ and $c^{(1,0)}(\rho )$ describe the order-$\alpha _{S}^{2}$
corrections, factorization scale ($\mu _{F}=\mu _{R}=\mu $) dependent and
independent, respectively.

To calculate the NLO corrections to the single spin 
asymmetry $A(s)$, we need to take into account the virtual 
${\cal O}(\alpha _{em}\alpha _{S}^{2})$ corrections to the Born process (\ref{3}) 
and the real gluon emission in the photon-gluon fusion:
\begin{equation}
\gamma (k_{\gamma })+g(k_{g})\rightarrow Q(p_{Q})+\overline{Q}(p_{\stackrel{
\_}{Q}})+g(p_{g}).  \label{4.1}
\end{equation}

The scale dependent term $c^{(1,1)}(\rho )$ which is the coefficient of $\ln 
\frac{\mu ^{2}}{m^{2}}$ can be expressed explicitly in terms of the Born
cross section using the renormalization group arguments  \cite{3}. 
Contribution of the scale independent cross secton, $c^{(1,0)}(\rho )$, near
the threshold can be obtained with help of the soft gluon resummation method 
\cite{15,16,17}. To the next-to leading logarithmic  accuracy, the
soft gluon contribution to the photon-gluon fusion can be written in a
factorized form as:
\begin{equation}
\hat{s}^{2}\frac{\mbox{d}^{2}\hat{\sigma}}{\mbox{d}\hat{t}_{1}\mbox{d}\hat{u}
_{1}}\left( \hat{s},\hat{t}_{1},\hat{u}_{1}\right) \approx B^{\rm{Born}
}\left( \hat{s},\hat{t}_{1},\hat{u}_{1}\right) \left\{ \delta \left( \hat{s}+
\hat{t}_{1}+\hat{u}_{1}\right) +\sum_{n=1}^{\infty }\left( \frac{\alpha
_{S}(\mu )}{\pi }\right) ^{n}K^{(n)}\left( \hat{s},\hat{t}_{1},\hat{u}
_{1}\right) \right\} ,  \label{nn.7}
\end{equation}
where $\hat{t}_{1}=\hat{t}-m^{2}$, $\hat{u}_{1}=\hat{u}-m^{2}$ and $B^{\rm{
Born}}\left( \hat{s},\hat{t}_{1},\hat{u}_{1}\right) $ describes the Born
level $\gamma g$ cross section:
\begin{equation}
B^{\rm{Born}}\left( \hat{s},\hat{t}_{1},\hat{u}_{1}\right)
=e_{Q}^{2}\alpha _{em}\alpha _{S}\left[ \frac{\hat{t}_{1}}{\hat{u}_{1}}+
\frac{\hat{u}_{1}}{\hat{t}_{1}}+\frac{4m^{2}\hat{s}}{\hat{t}_{1}\hat{u}_{1}}
\left( 1-\frac{m^{2}\hat{s}}{\hat{t}_{1}\hat{u}_{1}}\right) \right]
\label{n.8}
\end{equation}
At NLO, ${\cal O}(\alpha _{em}\alpha _{S}^{2})$, the soft gluon corrections
to NLL accuracy in the $\overline{\mbox{MS}}$ factorization scheme are (cf. Ref. \cite{17}):
\begin{eqnarray}
K^{(1)}\left( \hat{s},\hat{t}_{1},\hat{u}_{1}\right) &=&2C_{A}\left[ \frac{
\ln \left( \hat{s}_{4}/m^{2}\right) }{\hat{s}_{4}}\right] _{+}+ 
\left[ \frac{1}{\hat{s}_{4}}\right] _{+}\left\{ C_{A}\left( \ln \left( 
\frac{\hat{t}_{1}}{\hat{u}_{1}}\right) +\mbox{Re} L_{\beta }-\ln \left( 
\frac{\mu ^{2}}{m^{2}}\right) \right) -2C_{F}\left( \mbox{Re} L_{\beta
}+1\right) \right\} +  \nonumber \\
&&\delta \left( \hat{s}_{4}\right) C_{A}\ln \left( \frac{-\hat{u}_{1}}{m^{2}}
\right) \ln \left( \frac{\mu ^{2}}{m^{2}}\right),  \label{n.9}
\end{eqnarray}
where $C_{A}=N_{c}=3$, $C_{F}=\frac{N_{c}^{2}-1}{2N_{c}}=\frac{4}{3}$, 
$\hat{s}_{4}=\hat{s}+\hat{t}_{1}+\hat{u}_{1}$ and $\beta =\sqrt{1-\rho }$. 
The function $L_{\beta }$ and the plus-distribution in (\ref{n.9}) are defined as:
\begin{equation}
L_{\beta }=\frac{1-2m^{2}/\hat{s}}{ \beta} \left[ \ln \left( \frac{
1-\beta } { 1+\beta } \right) +\rm{i}\pi \right] ,  \label{nn.11}
\end{equation}
\begin{equation}
\left[ \frac{\ln ^{l}\left( \hat{s}_{4}/m^{2}\right) }{\hat{s}_{4}}\right]
_{+}=\lim_{\Delta \rightarrow 0}\left\{ \frac{\ln ^{l}\left( \hat{s}
_{4}/m^{2}\right) }{\hat{s}_{4}}\theta \left( \hat{s}_{4}-\Delta \right) +
\frac{1}{l+1}\ln ^{l+1}\left( \frac{\Delta }{m^{2}}\right) \delta \left( 
\hat{s}_{4}\right) \right\} .  \label{n.10}
\end{equation}

Eq.(\ref{n.9}) describes very well the threshold behavior of the
exact partonic cross sections $c^{(1,0)}(\rho )$ and $c^{(1,1)}(\rho )$ and 
gives, at NLO, practically whole contribution to the 
unpolarized bottom production in photon-hadron reactions
at fixed target energies, $E_{\gamma }\lesssim 1$ TeV.  

All terms in the r.h.s. of Eq.(\ref{n.9}) originate from the so-called
collinear $\vec{p}_{g,T}\rightarrow 0$, and soft, $\vec{p}_{g}\rightarrow 0$
, components of cross section.
Since the azimuthal angle $\varphi $ is the same for both $\gamma g$ and $Q
\overline{Q}$ centre of mass systems in the collinear and soft limits, 
Eq.(\ref{nn.7}) can be generalized to the
spin-dependent case substituting the spin-averaged Born cross section  
by the $\varphi $-dependent one: 
$B^{\rm{Born}}\left( \hat{s},\hat{t}_{1},\hat{u}_{1}\right)\longrightarrow
B^{\rm{Born}}\left( \hat{s},\hat{t}_{1},\hat{u}_{1},\varphi \right)$.

The results of our calculations of the single spin asymmetry $A(s)$ at NLO 
to NLL accuracy in $c$- and $b$-quark production are presented by dashed 
line in Fig.2. The details of calculations as well as the higher order predictions 
for $A(s)$ will be given in \cite{we2}.
\begin{figure}
\begin{center}
\begin{tabular}{cc}
\mbox{\epsfig{file=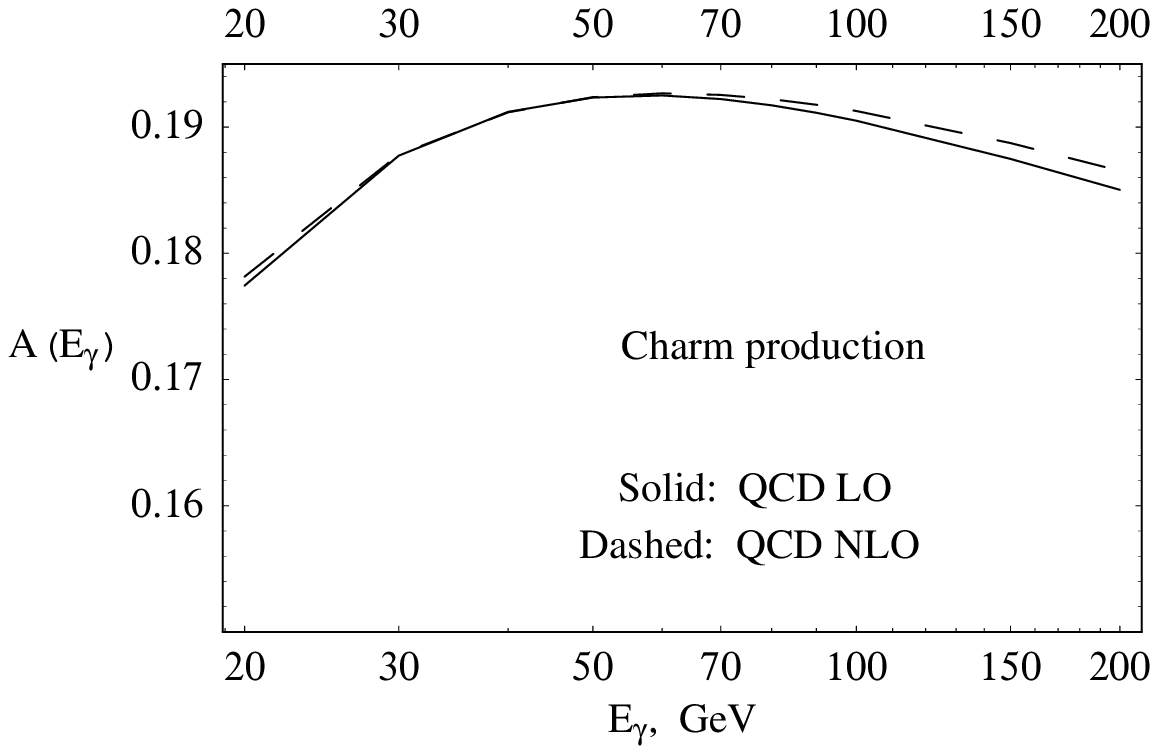,width=250pt}}
&\mbox{\epsfig{file=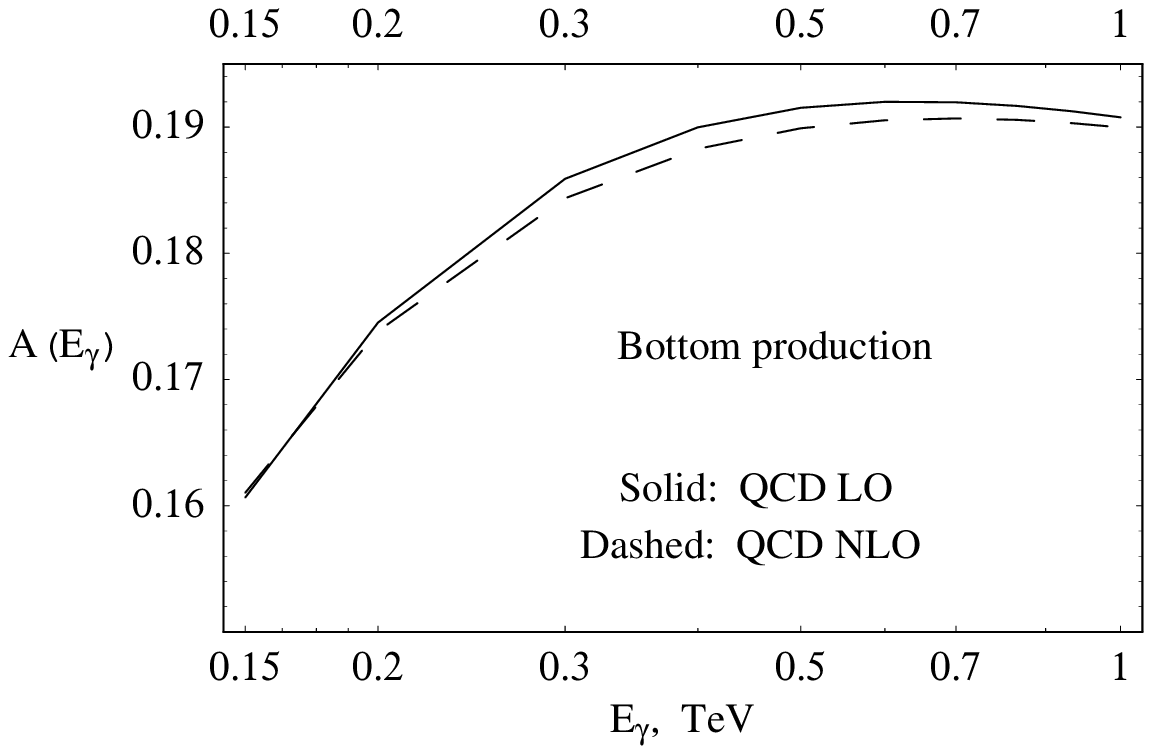,width=250pt}}\\
\end{tabular}
\caption{\small Single spin asymmetry, $A(E_{\gamma})$, in $c$- and $b$-quark
production at  LO (solid curve) and at NLO to NLL accuracy (dashed curve).}
\end{center}
\end{figure}

One can see from Fig.2 that the NLO NLL and Born predictions for $A(s)$ coincide 
with each other with accuracy better than 2\%. 
We have verified that the azimuthal asymmetry is independent 
(to within few percent) of theoretical uncertainties in the QCD input parameters 
($m_{Q} $, $\mu_{R}$, $\mu _{F}$ and $\Lambda _{QCD}$) at NLO too.

\section{Conclusion}

\noindent   Our analysis shows that the NLO corrections practically do not affect the 
Born  predictions for the single spin asymmetry in heavy flavor production by linearly
polarized photons at fixed target energies. 
So, the quantity  $A(s)$  is an observable quantitatively well defined, rapidly
convergent and insensitive to nonperturbative contributions. Measurements of
the azimuthal asymmetry in bottom production would be a good test of the
conventional parton model based on pQCD.

\acknowledgements 

\noindent The research described in this paper was carried out in collaboration 
with A. Capella and A.B. Kaidalov. Author would  like to thank A.V. Efremov for
useful discussion. I also wish to thank Organizing Committee of XV ISHEPP for 
invitation and financial support and, once more, E.B. Plekhanov for warm hospitality. 
This work was supported in part by the grant NATO PST.CLG.977275.

\end{document}